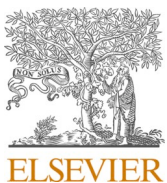
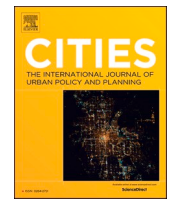

# Street review: A participatory AI-based framework for assessing streetscape inclusivity

Rashid Mushkani [a,b,*], Shin Koseki [a,b]

[a] *Université de Montréal, Canada*
[b] *Mila–Quebec AI Institute, Canada*

A B S T R A C T

Urban centers undergo social, demographic, and cultural changes that shape public street use and require systematic evaluation of public spaces. This study presents Street Review, a mixed-methods approach that combines participatory research with AI-based analysis to assess streetscape inclusivity. In Montréal, Canada, 28 residents participated in semi-directed interviews and image evaluations, supported by the analysis of approximately 45,000 street-view images from Mapillary. The approach produced visual analytics, such as heatmaps, to correlate subjective user ratings with physical attributes like sidewalk, maintenance, greenery, and seating. Findings reveal variations in perceptions of inclusivity and accessibility across demographic groups, demonstrating that incorporating diverse user feedback can enhance machine learning models through careful data-labeling and co-production strategies. The Street Review framework offers a systematic method for urban planners and policy analysts to inform planning, policy development, and management of public streets.

## 1. Introduction

Urban environments continue to undergo changes in demographic composition and cultural norms due to shifting migration patterns, economic developments, and mobility preferences (Anttiroiko & De Jong, 2020; Broderick, 2022; Youngbloom et al., 2023). City streets, sidewalks, and public areas often serve as primary interaction points among diverse user groups, including residents, commuters, and visitors (Gehl, 2011). These spaces carry social, economic, and cultural significance that influences navigation and user experience (Mitrašinović & Mehta, 2021).

Municipal governments and planning agencies recognize the importance of inclusive public spaces but face challenges in operationalizing inclusivity (Anttiroiko & De Jong, 2020). Traditional approaches may draw on universal design principles intended to accommodate a broad range of users, but these frameworks often take a one-size-fits-all approach that prioritizes physical accessibility over the social and cultural dimensions of public space use (Low, 2020). In multicultural cities, where multiple languages, cultures, and religious practices converge, these complexities become particularly evident (Fan et al., 2023; Litman, 2025; Salgado et al., 2021; Youngbloom et al., 2023).

Research on inclusive design has provided valuable insights, but few methods combine qualitative depth with quantitative scale to understand inclusivity in urban contexts (Anttiroiko & De Jong, 2020; Mehta, 2019; Zamanifard et al., 2019). Ethnographic research and interviews offer detailed perspectives on lived experience, while computer vision and machine learning enable assessments at larger scales (Ibrahim et al., 2020). However, large-scale computational approaches often overlook intersectional dimensions (Zhu et al., 2025). This gap calls for integrated models that merge qualitative and quantitative methodologies.

The Street Review framework addresses this gap by combining participatory methods with AI-based image analysis. The framework integrates co-production principles and intersectional theory with computer-based tools to establish a replicable protocol for measuring perceived inclusivity at both granular and broader urban scales. This approach leverages qualitative insights and automated image assessment to support analysis, modeling, and management of urban systems (Batty, 2018; Danish et al., 2025; Engin et al., 2020).

Municipal authorities and nonprofit organizations require methods to assess whether urban street environments serve diverse demographic groups effectively (Anttiroiko & De Jong, 2020; Jian et al., 2020; McKercher, 2020). While physical design features, such as sidewalks or bike lanes, may adhere to established guidelines, subtle social cues, cultural markers, and factors related to safety and belonging can differ among user groups (Fan et al., 2023; Rwiza, 2019). This research addresses the following questions:

1. How do individuals from different backgrounds perceive and experience inclusion or exclusion in Montréal's streetscapes?
2. Which urban design features and socio-demographic factors influence perceived inclusivity, and how do these factors vary across age, gender, ethnicity, religious identity, and ability?
3. Can a co-produced, AI-driven tool offer reliable citywide assessments of inclusivity that incorporate diverse user perspectives?
4. What guidelines can support urban planners and policymakers in designing streets that acknowledge the needs of different user groups?

* Corresponding author at: School of Urbanism and Landscape Architecture, Faculty of Environmental Studies, University of Montreal, 2940 Côte-Saint-Catherine, Montreal, Quebec, H3S 2C2, Canada.
*E-mail address:* rashid.ahmad.mushkani@umontreal.ca (R. Mushkani).

The paper is organized as follows. The introduction contextualizes the study by discussing the need for inclusive streets and outlining the main research questions. A review of relevant literature follows, examining concepts such as intersectionality, and the role of AI in urban analysis. The methodology section details the multi-stage design, sampling, and data-collection processes. Next, the application of the Street Review method in Montréal is presented, highlighting correlations and divergences in perceptions of inclusivity. A subsequent discussion interprets the results in light of theories of urban design, co-production, and AI ethics. The conclusion summarizes contributions and offers recommendations for future research and practice.

## 2. Literature review

Public spaces in urban contexts serve multiple functions, including civic engagement, economic activities, and cultural exchange (Gehl, 2011). Streets and sidewalks often constitute the core of public life, facilitating interactions that shape collective experiences (Whyte, 2021). However, individuals and groups encounter these spaces differently. Research in social geography and urban sociology illustrates that access to public space is influenced by factors such as income, race, ethnicity, gender, physical ability, and cultural affiliations (Armstrong & Greene, 2022; Costanza-Chock, 2020; Sadeghi & Jangjoo, 2022).

Intersectionality offers a framework for understanding how multiple forms of discrimination or privilege converge to influence people's interactions with urban environments (Crenshaw, 1989). Urban planners and policymakers increasingly acknowledge that standardized approaches to street design may overlook diverse needs (Dmowska & Stepinski, 2018; Lawton Smith, 2023; Low, 2020). Intersectional perspectives suggest that measures aimed at improving accessibility for individuals with physical impairments may not address the needs of other populations, such as religious minorities or LGBTQIA+ groups, who encounter distinct social barriers (Rinaldi et al., 2020; Stark & Meschik, 2018; Talen, 2012).

In response to these complexities, participatory planning and co-production frameworks have emerged as strategies for involving community members in decision-making processes (McKercher, 2020; Rinaldi et al., 2020). These approaches propose that user insights, lived experiences, and cultural knowledge can inform policy-making beyond conventional expert-driven models. Participatory methods include public workshops, focus groups, citizen advisory committees, and iterative co-design processes (Asaro, 2000; Creswell & Creswell, 2022; Fors et al., 2021).

Participatory processes can reveal dynamics such as how certain groups perceive safety in spaces that others consider neutral (Tandogan & Ilhan, 2016). Research in street design may examine factors ranging from lighting and bench placement to symbolic displays of cultural identity (Biljecki et al., 2023). Such methods acknowledge that users hold expertise in their environments, and this expertise is critical in shaping inclusive outcomes (Fischer, 2000; Gibbons et al., 1994).

Advances in computer vision and machine learning have introduced automated methods for analyzing streetscapes on a large scale (Cheliotis, 2020; Ibrahim et al., 2020). Platforms like Google Street View and Mapillary provide geotagged images that researchers can analyze to identify features such as greenery, building heights, or façade conditions, characterizing urban form (Danish et al., 2025; Zhu et al., 2025). These computational tools enable the assessment of thousands of images, offering decision-makers cost-effective audits of street conditions (Huang et al., 2023; Ibrahim et al., 2020; Liu et al., 2017).

Efforts such as Place Pulse, StreetScore, and Project Sidewalk have pioneered the use of technology to assess urban environments. Place Pulse employs crowdsourced surveys to generate datasets of subjective perceptions regarding safety, liveliness, and beauty in streetscapes (Dubey et al., 2016). Building on this, StreetScore applies computer vision and machine learning to predict safety perceptions at scale, translating these human judgments into automated evaluations (Naik et al., 2014). Similarly, Project Sidewalk uses a gamified interface for virtual audits of sidewalk conditions, identifying barriers for individuals with mobility impairments (Saha et al., 2019). While these initiatives demonstrate the promise of merging human insight with computational analysis, they often rely on predefined criteria and may embed biases (Angwin et al., 2022). The proposed Street Review method extends these approaches by embedding a co-production framework that integrates diverse stakeholder perspectives into supervised machine learning models. This iterative refinement of evaluation criteria seeks to capture a broader range of intersectional experiences, addressing limitations identified in earlier projects.

Recent studies have integrated automated image analysis, open-source street-level imagery, and resident feedback to evaluate urban environments. Kang et al. (2023) demonstrated that GeoAI-based safety predictions diverged from neighborhood survey responses, underscoring the need to align computational outputs with lived experience. Yang et al. (2025) developed tools for computing visual indicators but left questions of data quality and representativeness open.

Other studies have used large-scale visual datasets and crowdsourced ratings to assess perceptions of urban form. Ogawa et al. (2024) trained a deep-learning model on 8.8 million ratings to predict 22 attributes of urban form. Cui et al. (2023) identified systematic gender-based differences in safety perceptions. Ito et al. (2024), in a review of 393 studies, highlighted limitations in spatial scope, label quality, and causal inference. The Street Review approach addresses these challenges by producing ground-truth labels from qualitative interviews and focus group image evaluations, emphasizing intersectional and context-specific perspectives. These inputs inform a supervised multi-output regression model that does not rely on anonymous or single-dimensional crowdsourced data.

Discussions of AI-based systems caution that such methods can perpetuate social biases if training data or labeling processes do not encompass diverse perspectives (Barocas et al., 2022; Buolamwini & Gebru, 2018). For example, an algorithm might underestimate the significance of communal seating if that feature is important primarily to a specific user group underrepresented in the dataset (Malekzadeh et al., 2025). Similarly, a model might misinterpret religious markers without training to recognize their role in shaping perceptions of inclusivity (Wang et al., 2022). These considerations underline the need for balanced datasets and transparent, accountable modeling practices (Mehrabi et al., 2021).

Researchers advocate for ethical AI frameworks that incorporate community feedback from the early stages of system design (Buolamwini & Gebru, 2018). Co-production approaches in AI development engage diverse stakeholders in data collection, labeling, model refinement, and result interpretation (Mushkani, Berard, Cohen, et al., 2025). This pluralistic paradigm seeks to incorporate diverse social and cultural values into algorithmic outputs (Varanasi & Goyal, 2023).

Bridging intersectionality with co-production provides a framework for designing AI systems that are both contextually aware and reflective of diverse social realities (Crenshaw, 1989). This integration involves iterative feedback loops, where community members review model results, identify inaccuracies, and propose adjustments. In the context of street design, such methods can prevent the creation of generic "inclusivity" maps that fail to reflect the nuanced experiences of diverse groups (Johnson & Miles, 2014; Lee, 2022; Roberson, 2022; Stark & Meschik, 2018).

This study bridges gaps at the intersection of three critical areas: intersectional urban studies, participatory planning, and AI-driven methods. While inclusive streets are widely recognized as vital, there is a lack of methodologies that integrate qualitative, user-driven insights with scalable machine learning tools (Anttiroiko & De Jong, 2020; Huang et al., 2023; Zhu et al., 2025). While studies combining AI with urban design often overlook intersectional experiences, participatory approaches frequently face scalability challenges.

To address these gaps, this paper proposes an integrated framework (Street Review) that uses co-production to embed user perspectives into supervised machine learning models. By integrating qualitative interviews, focus groups, image ratings, and large-scale analysis, the framework provides actionable insights for planners and policymakers assessing urban street inclusivity.

## 3. Methodology

We employed a multi-phase design to gather diverse data on urban streetscapes in Montréal, between mid-2023 and late 2024. Our primary goal was to establish a replicable workflow for measuring perceived inclusivity in streets. This workflow comprised: (1) semi-directed interviews, (2) individual and group-based image rating, (3) image collection and data preparation, (4) AI model training and validation, and (5) inference and heatmap generation.

### 3.1. Participant recruitment and interviews

We contacted more than 100 community organizations, local associations, and service providers in Montréal, aiming to recruit participants with varied religious backgrounds, ethnicities, ages, genders, socio-economic statuses, disability statuses, and sexual orientations (Creswell & Creswell, 2022; IRCGM, 2018). A total of 35 individuals expressed interest, out of which 28 took part in semi-directed interviews. We prioritized historically underrepresented or marginalized groups, including recent immigrants and individuals with limited mobility (Creswell & Creswell, 2022).

Fig. 1 depicts the distribution of self-declared diversity characteristics across various age brackets. Categories such as senior citizens, women, ethnic minorities, persons with disabilities, LGBTQ2+, and religious minorities are represented. This figure underscores our commitment to an inclusive and intersectional recruitment strategy that considers participants' multiple identities. While capturing a range of perspectives on urban streetscapes, we acknowledge that these identities do not solely determine how participants perceive their environments but rather provide multiple lenses for analysis (Benjamin, 2019; Crenshaw, 1989).

Each interview lasted 30 to 90 min. Participants viewed street-level images from different neighborhoods, spanning commercial zones, residential streets, mixed-use corridors, older districts, and newly developed areas. The interview format followed a semi-structured protocol organized around three core prompts: "When we say public space, what comes to mind?", "What are your favorite and least favorite public spaces in Montréal?", and "What qualities make a street or park work for you?" Follow-up questions probed themes such as accessibility, comfort, and exclusion, and were adapted in real time to reflect each participant's background and lived experience. This format enabled thematic consistency while allowing new concerns to emerge (Bryman, 2012).

Demographic variation in responses was notable. Participants with mobility impairments emphasized curb cuts, ramp access, and sidewalk maintenance. Elderly respondents highlighted lighting, vehicle speed, and signal clarity. Younger participants focused on flexible usage and visual interest, while LGBTQ2+ individuals referenced cues of acceptance and the presence of vibrant nighttime activity. These group-specific insights informed our thematic coding strategy and the selection of the four perceptual criteria used for model training: accessibility, aesthetics, practicality, and inclusivity (Section 3.3).

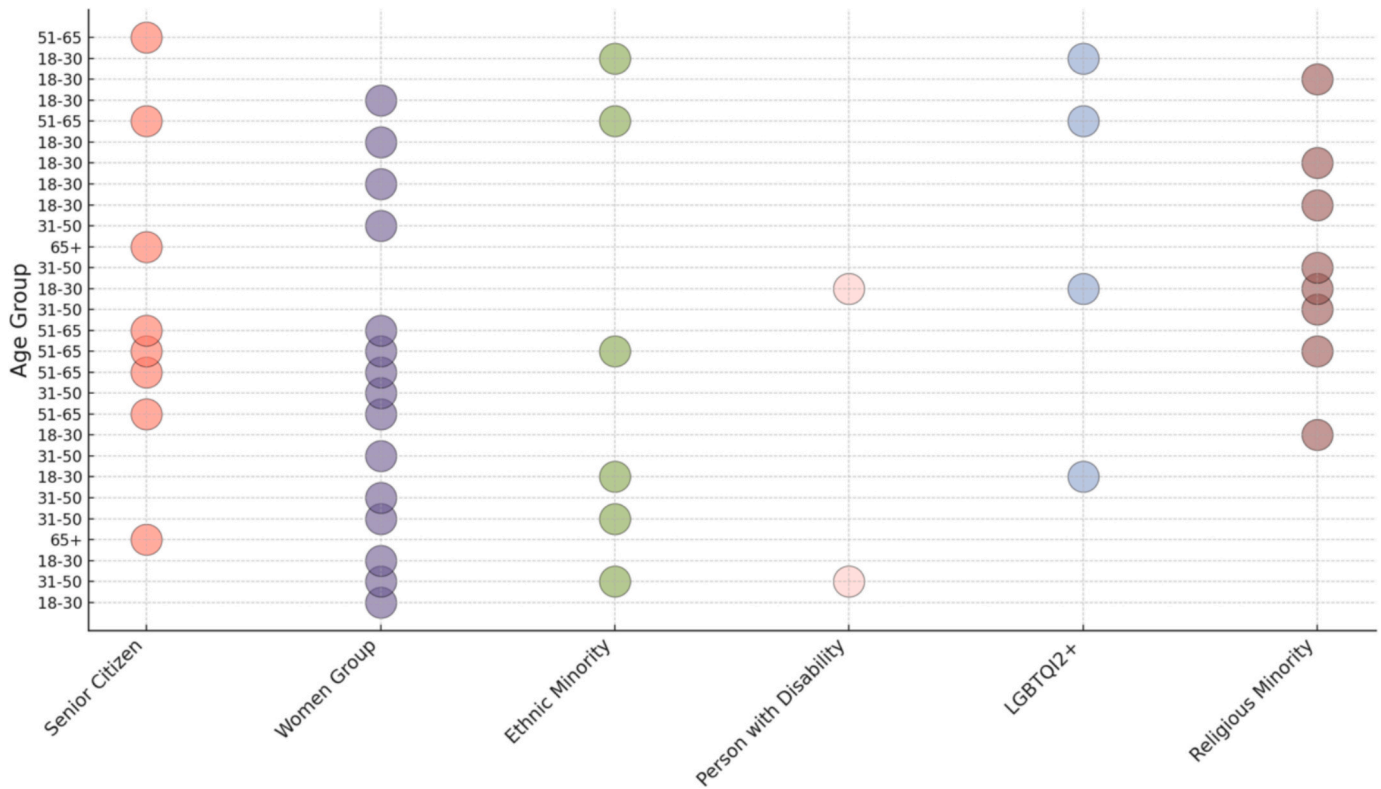

**Fig. 1.** Diversity representation by age group. Distribution of self-declared diversity characteristics across age categories, highlighting intersectional perspectives without assuming identity solely determines perception (Mushkani, Nayak, et al., 2025; Mushkani, Berard, & Kosek, 2025).

Ethical safeguards were in place throughout. Participants were informed of their right to skip any question or end the interview at any time. This approach encouraged open, self-directed dialogue and supported the inclusion of diverse perspectives (Bryman, 2012).

*3.2. Spatial scope and data collection*

We selected 20 street locations across Montréal as study sites. For each street, three data points were identified at distinct positions—head, center, and tail—yielding a total of 60 data points for focus group exercises. As part of the evaluation process, participants initially assessed each data point based on two representative images. To prepare the dataset for AI training and capture varied perspectives of the same location, we expanded the image collection to approximately 250 street-view images per data point, captured in 360-degree rotating frames, resulting in a local dataset of 15,000 images (60 data points × 250 images = 15,000) (Goodfellow et al., 2016).

Twelve participants scored each data point, represented by two images, on four perceptual criteria (total of 120 images). The scores were averaged within six demographic groups (LGBTQ2+, mobility-impaired, elderly female, elderly male, young female, and young male). Each point's scores were then assigned to all 250 frames captured at that location, as the frames represent the same scene from contiguous angles. As noted in the interviews, participants found that evaluating a space using a single image often omitted important context, while 360-degree images introduced visual distortion that made evaluations difficult. Therefore, we did not use 360-degree images for perceptual assessments; instead, each point was captured using two images from opposing angles (Ausin-Azofra et al., 2021; Hussain & Kwon, 2021).

Fig. 2 illustrates the geographic spread of the selected streets, reflecting socio-economic diversity and varying land-use patterns. This selection aimed to address spatial equity in the study of inclusivity (Low, 2020). The local dataset was complemented with approximately 45,000 geotagged images from the crowd-sourced Mapillary platform for city-wide analysis. These images predominantly represent main streets and may not cover all areas equally, leaving some streets underrepresented.

The integration of both locally collected and Mapillary images supports heatmap generation, model validation, and the analysis of spatial patterns of inclusivity and accessibility across diverse neighborhoods.

*3.3. Thematic analysis*

We transcribed and coded the interview audio recordings to identify recurring concepts such as accessibility, aesthetics, safety, community engagement, maintenance, and sense of belonging (Creswell & Creswell, 2022; Miles & Huberman, 2003). These categories formed an overarching framework of four perceptual criteria: accessibility, aesthetics, practicality, and inclusivity.

Table 1 presents the frequency counts of seven manually coded themes across four primary demographic groups. We interpreted these patterns as follows:

1. Accessibility (Theme 1) ranked first among participants with direct experience of disability. Forty-five percent of their coded statements addressed curb cuts, ramp access, sidewalk maintenance, or related features. One mobility-impaired participant whose mother uses a

**Table 1**
Frequency of themes across demographic groups based on interview data.

| Theme | Elderly (n = 13) 178 statements | Mobility-impaired (n = 2) 79 statements | Young adults (n = 8) 150 statements | LGBTQ2+ (n = 5) 88 statements |
| --- | --- | --- | --- | --- |
| Accessibility and safety | 67 (37.6 %) | 36 (45.6 %) | 21 (14.2 %) | 12 (13.6 %) |
| Inclusivity and sense of belonging | 36 (20.2 %) | 16 (20.3 %) | 15 (10.1 %) | 20 (22.7 %) |
| Functional design and utility | 26 (14.6 %) | 10 (12.7 %) | 23 (15.5 %) | 9 (10.2 %) |
| Aesthetic and maintenance | 22 (12.4 %) | 6 (7.6 %) | 42 (28.4 %) | 14 (15.9 %) |
| Management and responsibility | 8 (4.5 %) | 4 (5.1 %) | 8 (5.4 %) | 7 (8.0 %) |
| Community engagement | 9 (5.1 %) | 3 (3.8 %) | 27 (18.2 %) | 16 (18.2 %) |
| Historical significance and others | 10 (5.6 %) | 4 (5 %) | 12 (8.2 %) | 10 (11.2 %) |

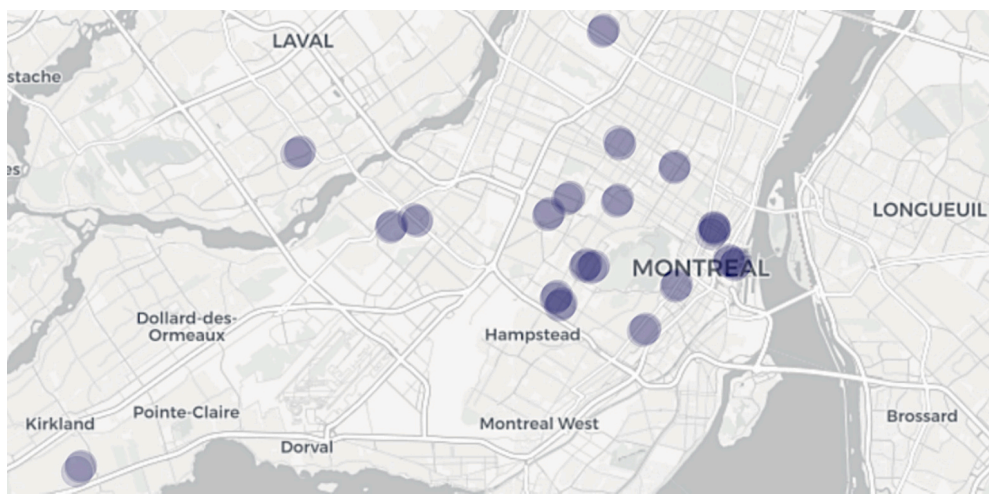

**Fig. 2.** Spatial distribution of study sites. Geographic spread of the 20 selected street locations across Montréal. The clustering patterns reflect a diverse representation of socio-economic contexts, land uses, urban densities, and historical periods. The base map data is sourced from OpenStreetMap (Mushkani & Koseki, 2025; Mushkani, Berard, & Kosek, 2025).

wheelchair summarized, "*Even a small flight of stairs means we have to turn around.*"
2. Safety, a component of Theme 1, was a primary concern for elderly participants. Thirty-one percent of their statements addressed lighting, vehicle speed, or signal clarity. One older woman noted, "*Not enough lights on the street… you must look twice before crossing.*" Safety was also linked to Theme 3: functional design and utility. One elderly participant commented, "*The sidewalks are this big… the bicycle paths and the car lanes are that big… there's no space for that anymore.*"
3. Aesthetic and maintenance concerns (Theme 4) were most frequently mentioned by younger adults. Twenty-eight percent of their statements centered on visual interest or place comfort. One participant noted, "*A space should invite you to stay, not only pass through.*"
4. Inclusivity and welcoming features (Theme 2) were emphasized by LGBTQ2+ participants. Twenty-two percent of their statements referenced cues such as multilingual signage and nighttime streets activities.

We originally defined four demographic groups: elderly adults, mobility-impaired individuals, young adults, and LGBTQ2+ persons. Because single-axis groupings can obscure intersectional attributes (for example, two mobility-impaired participants also identified as elderly and as women), we expanded the final street review model's demographic schema to six groups: LGBTQ2+, mobility-impaired, elderly female, elderly male, young female, and young male.

To validate the identified themes, we conducted a Latent Dirichlet Allocation (LDA) analysis on the interview transcripts (Blei et al., 2003). LDA identified latent topics within the data and allowed us to assess the degree of convergence or divergence between algorithmically derived topics and the initial thematic analysis results. Each topic was labeled by reviewing its top words and representative text segments, assigning a meaningful theme that matched or refined the manual codes.

We then built a co-occurrence matrix to capture how often pairs of these topics appeared together within the same document or text segment. This matrix was used to define a network, where nodes represent themes and edges indicate the strength of their co-occurrence. The network was visualized using a graph layout algorithm to highlight the relationships between themes. Fig. 3 presents the network of themes derived from participant interviews, showing the co-occurrence patterns among accessibility, safety, inclusivity, and aesthetics. This procedure aligned the manual coding with topic modeling outputs and established consistency across analytical methods. From the set of recurrent and frequently co-occurring themes, we derived four perceptual criteria—accessibility, aesthetics, practicality (Theme 3: functional design and utility), and inclusivity. These criteria structured both the image rating tasks in focus groups and the training of the AI model (Creswell & Creswell, 2022).

*3.4. Image rating and ranking*

We invited 28 individuals from the original interview pool to participate in focus groups, and 12 accepted our invitation to assist. Each group session lasted approximately 3 h and included four stages:

1. Individual rating: Participants assigned scores to 120 images from 20 streets. Each street was represented by 3 vantage points, with 2 images per vantage point. Ratings were based on practicality, aesthetics, accessibility, and inclusivity, using a four-point scale (1 =

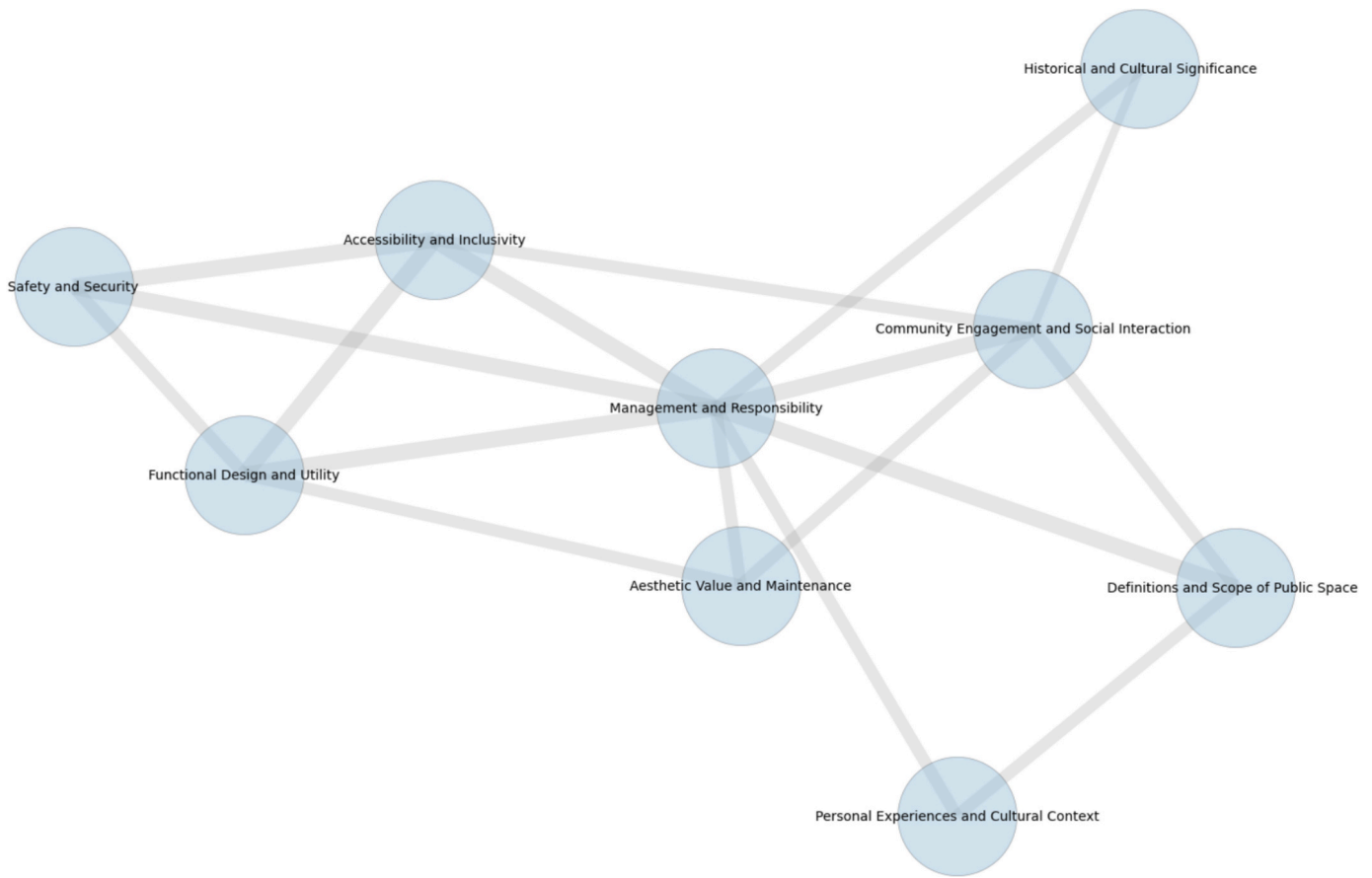

**Fig. 3.** Network of interview transcription themes. Thematic relationships derived from participant interviews, highlighting interconnected concepts such as accessibility, safety, inclusivity, and aesthetics.

**Table 2**
Rating scores.

| Dimension | Score 1 | Score 2 | Score 3 | Score 4 |
|---|---|---|---|---|
| Inclusivity | Not inclusive or welcoming | Some inclusivity measures present | Broadly welcoming and inclusive | Fully inclusive and welcoming to all |
| Aesthetics | Poor design and minimal greenery | Basic design with limited greenery | Appealing design with abundant greenery | Highly attractive with rich, diverse greenery |
| Practicality | Non-functional and poorly maintained | Barely functional, maintenance lacking | Adequately functional with regular upkeep | Highly functional with proactive maintenance |
| Accessibility | Inaccessible | Limited accessibility | Generally accessible, some difficult areas | Fully accessible for all users |

poor, 4 = excellent). See Table 2 for detailed rating information (Mushkani & Koseki, 2025; Mushkani, Berard, & Kosek, 2025).
2. Group discussion: Participants discussed their rationale for specific scores, identifying consensus or disagreement.
3. Collective rating: Groups reconciled differing views, producing a shared rating for each image.
4. Ranking task: Participants selected three images as most inclusive, three as least inclusive, and a middle group.

Fig. 4 provides examples of the diverse neighborhoods included in these rating exercises. Through these discussions, we collected qualitative insights on how amenities, seating, or signage might influence perceived inclusivity (Creswell & Creswell, 2022).

### 3.5. Diversity street selection

We designed a street diversity selection matrix to ensure variation in factors such as affordances (activities and amenities), greenery, space-to-user relationships, density, socio-economic status, urbanization spectrum, historical context, and land use (Talen, 2012; Ye, 2019). Fig. 5 illustrates the distribution of sampled streets across various urban characteristics—such as density, socio-economic status, greenery, and historical context—using horizontal bars to indicate their relative frequencies. The sampling approach was based on ensuring at least one street per category. For example, when selecting streets based on density, the goal was to include at least one from low-, medium-, and high-density areas. The final distribution, however, was not perfectly balanced; while all categories were represented, some were more prevalent than others.

### 3.6. Machine learning pipeline

We developed a multi-stage machine learning pipeline to predict practicality, aesthetics, accessibility, and inclusivity scores from street-view images. The pipeline consists of the following components:

1. Semantic segmentation: We employed the SegFormer-B5 model, a transformer-based semantic segmentation model with 82 million parameters (Xie et al., 2021). The model was fine-tuned on the CityScapes dataset at a resolution of 1024 × 1024 to classify pixels into predefined categories such as sidewalks, buildings, vegetation, and signage (Cordts et al., 2016). The encoder-decoder architecture of SegFormer ensures efficient high-resolution image processing while maintaining spatial consistency, as shown in Fig. 6 (Xie et al., 2021).
2. Feature extraction: Building on the pixel-wise segmentation masks from SegFormer, we derived a 12-dimensional feature vector for each pixel by preserving both its color intensities (R, G, B) and segmentation confidence scores for the most relevant classes. Specifically, we retained confidence scores for sidewalk, building, wall, fence, pole, traffic light, traffic sign, vegetation, and terrain, while discarding those for sky, vehicles, persons, road, motorcycles, and bicycles. As illustrated in Fig. 7, this process yields a pixel-level representation that captures subtle semantic and color cues critical for streetscape evaluation, ensuring minimal loss of fine-grained information (Chen et al., 2024; Goodfellow et al., 2016; Huang et al., 2023; Naik et al., 2014).

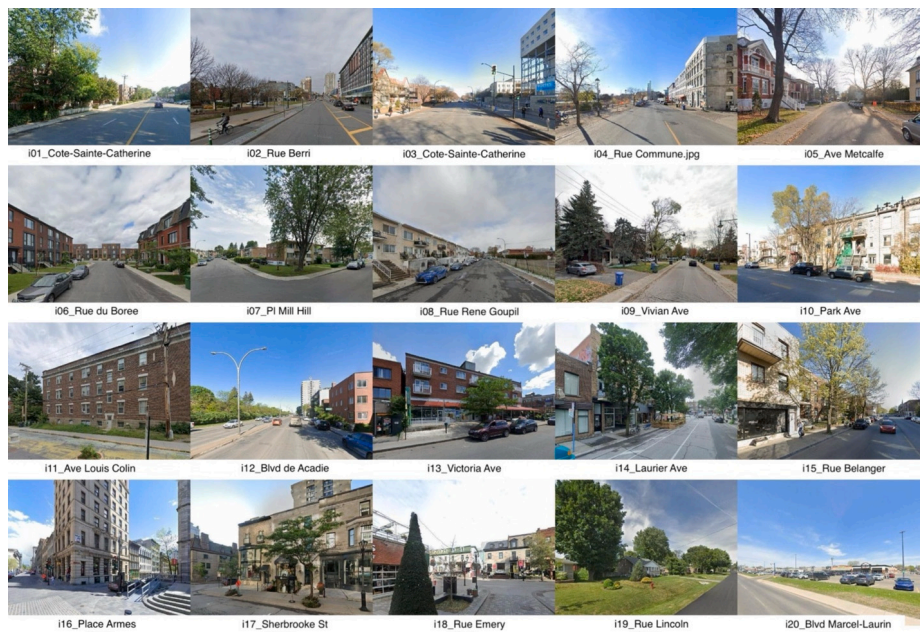

**Fig. 4.** A representative selection of 20 street-level images, each sampled from a dataset containing 250 image frames per point, captured in a 360-degree rotation. The images illustrate variations in land use, greenery, pedestrian amenities, and overall streetscape character across diverse neighborhoods, with three sampled points per street (Mushkani, Berard, Ammar, et al., 2025; Mushkani & Koseki, 2025; Mushkani, Berard, & Kosek, 2025).

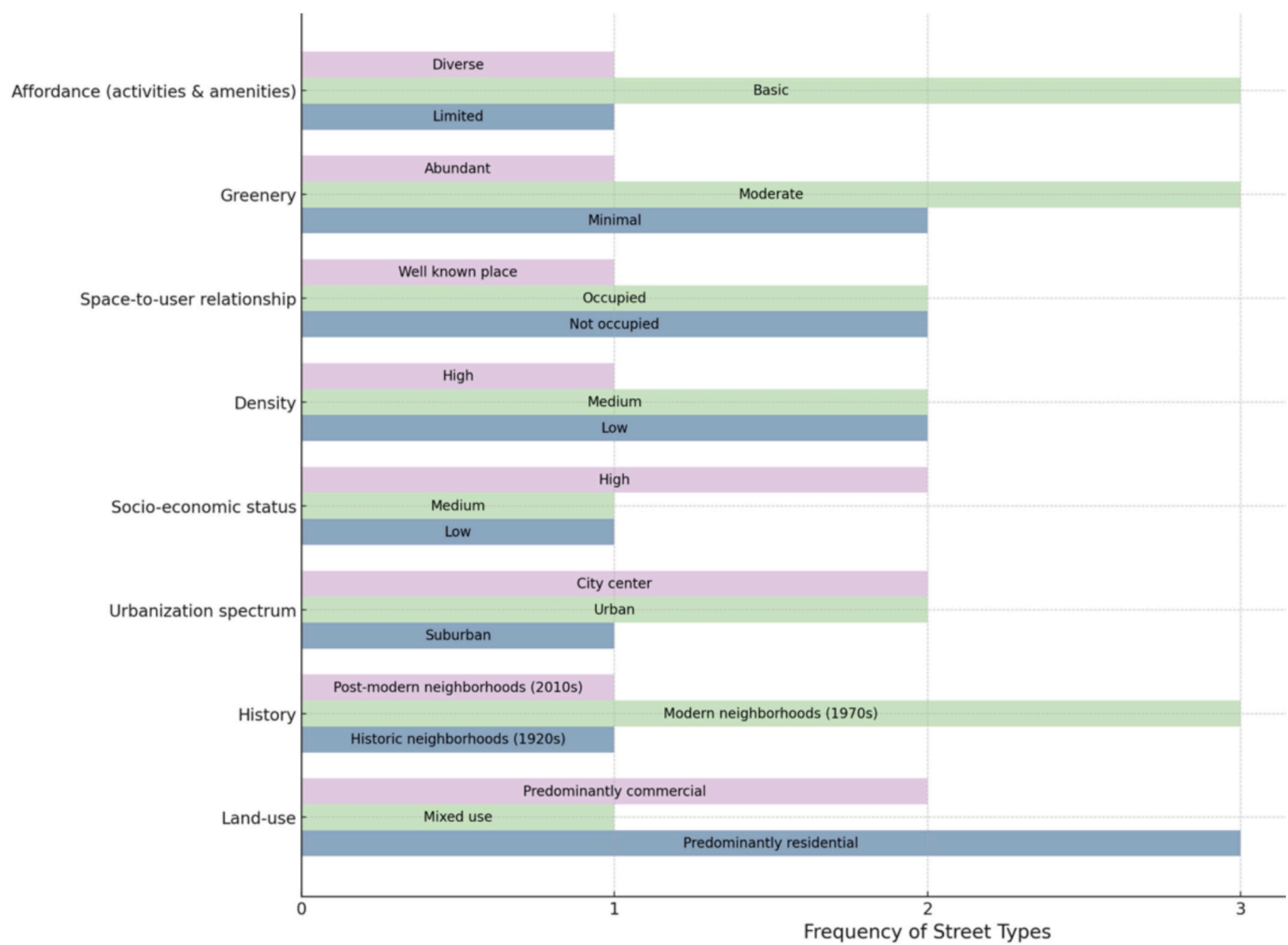

**Fig. 5.** Distribution of sampled streets across a spectrum of socio-spatial attributes, depicted through horizontal bars indicating the relative frequency of each street category (Mushkani, Berard, & Kosek, 2025).

3. Street review model: We employ a custom multi-layer perceptron (MLP) with six attention heads and 11 fully connected layers (Bishop, 2006; Vaswani et al., 2023), specifically designed to process pixel-level features without relying on aggregation or one-hot encodings. Instead of using pooled image summaries, the model processes sequences of 12-dimensional vectors (one per pixel) applying normalization followed by multi-head attention to capture pixel-level relationships. Mean pooling is applied only after attention outputs are computed. The model operates as a supervised multi-output regression system, predicting 28 scores: four criteria for each of six identity marker groups-LGBTQIA2+, individuals with disabilities, elderly women, elderly men, young men, and young women (24 outputs)-along with four group-level ratings. Despite its fine-grained input, the attention-based MLP remains computationally efficient, with fewer than one million trainable parameters. The multi-head attention mechanism enables the model to capture complex relationships among pixels and semantic classes (Brigato & Iocchi, 2020; Goodfellow et al., 2016), facilitating alignment with participant-defined ratings. Fig. 6 illustrates the architecture in detail, highlighting how the attention heads and fully connected layers jointly leverage pixel-level features to generate predictions across diverse evaluation criteria.

The model was trained on four NVIDIA V100-16GB GPU for 12 h, using mean squared error (MSE) as the loss function to optimize predictions against participant-assigned scores. It achieved $R^2$ scores of 0.91 on the validation set and 0.89 on the test set, indicating strong predictive accuracy. The coefficient of determination ($R^2$) quantifies the proportion of variance in the dependent variable explained by the model, with values near 1 signifying robust performance. Following training, inference was performed on a dataset of 45,000 street-view images, with feature extraction and prediction generation completed over two days, enabling a comprehensive evaluation of streetscapes at scale (Goodfellow et al., 2016).

The labeled data were divided into training, validation, and test sets (70 %, 15 %, 15 %) using a stratified split across evenly distributed vantage points to prevent data leakage (Bishop, 2006). The dataset comprises 250 locally collected images per data point, with 60 data points totaling 15,000 images, each paired with participant feedback to establish the initial ground truth.

Permutation importance analysis, illustrated in Fig. 8, reveals the relative contribution of streetscape features to the model's predictions of inclusivity, accessibility, practicality, and aesthetics (Molnar, 2025). We used this method to identify which visual features most strongly influenced the model's outputs and to validate whether these align with participant priorities. Permutation importance was calculated by randomly shuffling each feature 100 times and recording the resulting reduction in $R^2$ compared to the baseline model. Final scores represent the average $R^2$ drop across these shuffles, while standard errors indicate the stability of each estimate. Sidewalk coverage and building frontages emerged as the most significant predictors, followed by walls, reflecting participants' emphasis on boundary structures for walkability and safety. In contrast, fences and vegetation exhibited lower importance. Although focus group discussions indicated strong emotional responses to greenery, the model's low weighting of vegetation suggests difficulty in capturing subjective perceptions through simple proxies like green pixel proportion.

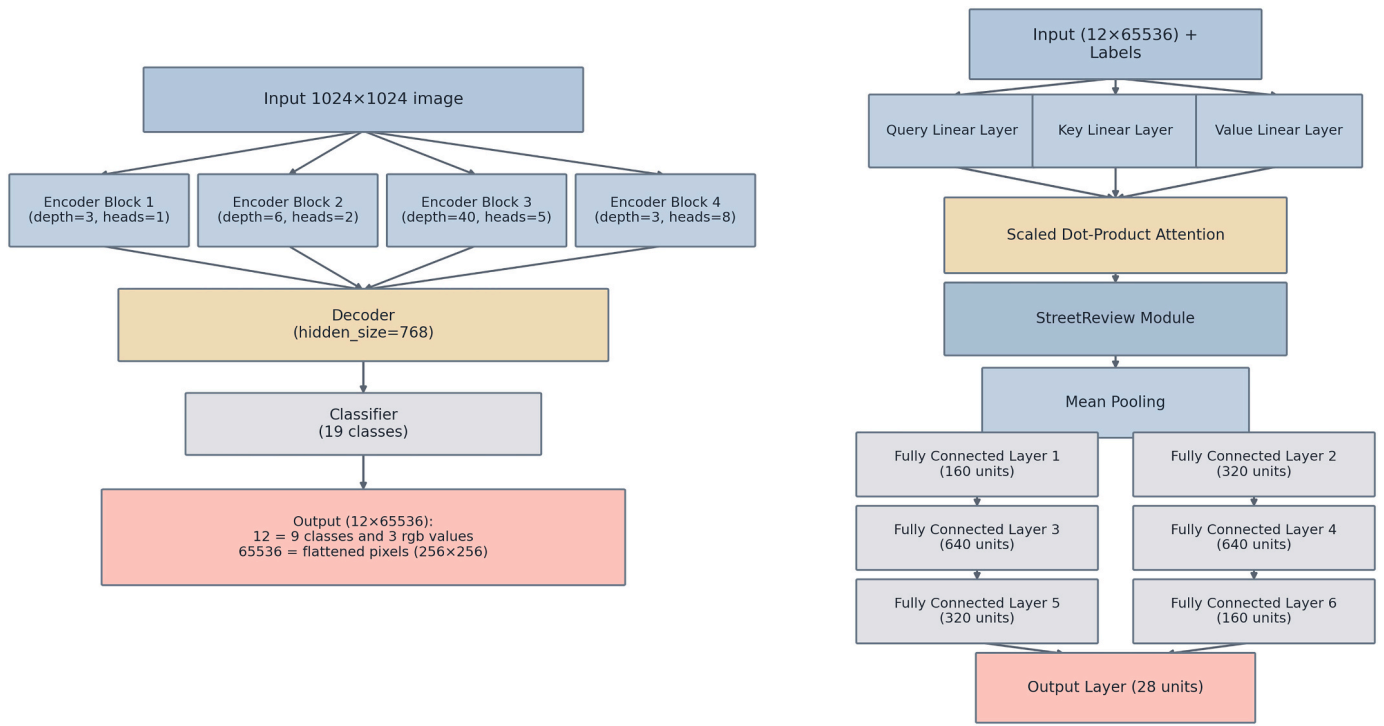

**Fig. 6.** Left: Overview of the Nvidia Segformer model, showing how encoder blocks and a decoder process high-resolution images to classify streetscape elements. Right: Structure of the 11-layer Street Review model, depicting how scaled dot-product attention and fully connected layers translate extracted features into inclusivity scores.

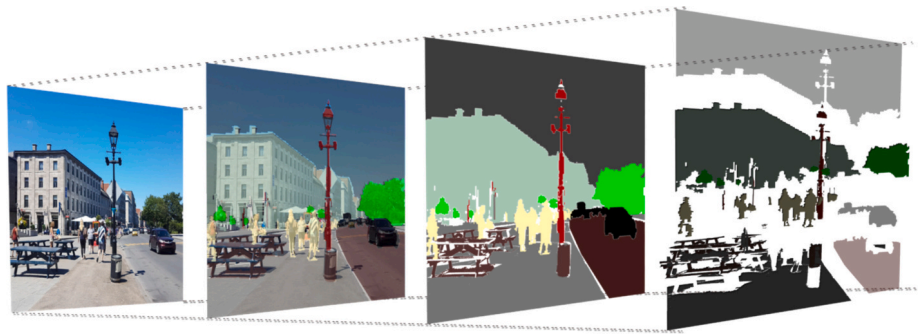

**Fig. 7.** Illustration of multi-stage processing. The process starts with raw street-view images, followed by semantic segmentation that assigns class labels to each pixel. In the final stage, only selected classes relevant to inclusivity evaluation are retained, while irrelevant ones such as cars, sky, and asphalt are dropped (shown in light gray). Location: Old port of Montreal.

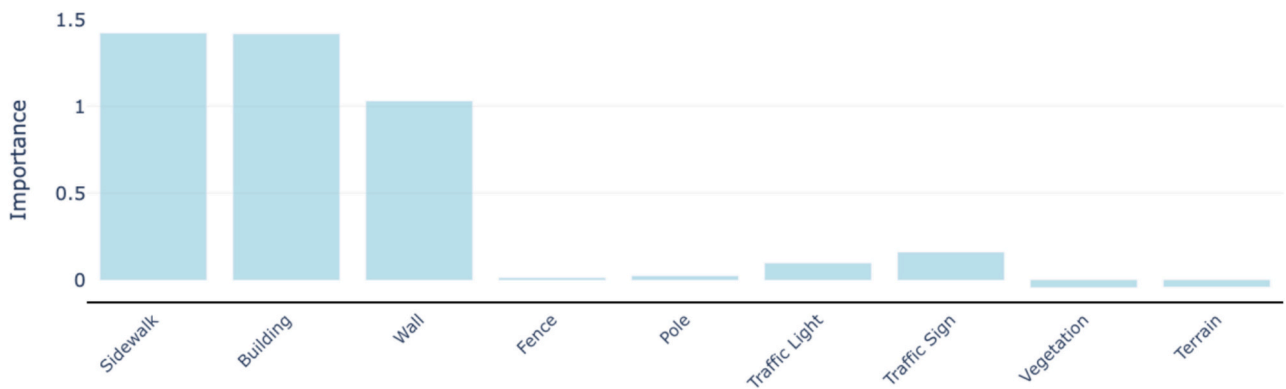

**Fig. 8.** Permutation importance. Relative contributions of various streetscape elements to the model's predictions of inclusivity, accessibility, practicality, and aesthetics as determined by permutation importance analysis.

Fig. 9 compares AI outputs with participant ratings, highlighting areas of effective performance and persistent differences across demographic groups or specific evaluation criteria (Goodfellow et al., 2016). To further evaluate model reliability, we assessed performance across all 28 group-criterion combinations. Fig. 10 presents the $R^2$ scores for each of these, illustrating how predictive accuracy varies by both demographic subgroup and evaluation criterion. This breakdown supports subgroup-level validation and shows where performance remains consistent or diverges across user identities and evaluation domains.

After confirming satisfactory model performance on the test set, we applied the trained model to the entire 45,000-image citywide dataset. We aggregated the resulting scores—practicality, aesthetics, accessibility, and inclusivity—at the street-segment level and generated heatmaps to visualize spatial patterns of inclusivity across Montréal. Additional demographic-specific layers were created by applying weights from participant subgroups, indicating variations in perceived inclusivity among different user identities (Goodfellow et al., 2016).

By integrating thematic analysis, focus group rating exercises, and a supervised machine learning pipeline, we created a framework for quantifying and visualizing perceived inclusivity in Montréal's streetscapes. This framework, shown in Fig. 11, combines community input, image ratings, and model training to predict and map scores across four evaluation criteria using street-level imagery. Our approach underscores the importance of incorporating diverse perspectives into urban analysis and planning.

## 4. Findings

This section synthesizes results from the participatory research modules and the machine learning analysis, focusing on correlations among the four evaluation criteria (Inclusivity, Accessibility, Practicality, and Aesthetics), demographic variations, and citywide inclusivity patterns in the model's predictions. The findings also address limitations of large-scale street-view image datasets and illustrate how demographic-specific weights can refine neighborhood assessments (Danish et al., 2025; Huang et al., 2023).

### 4.1. Citizen assessments

Fig. 12 presents the ratings assigned by 12 focus group participants to 60 data points, each corresponding to images of selected Montréal streets. Across all streets, the Group Accessibility category received an average score of 2.12, Group Inclusivity 2.06, Group Practicality 2.39, and Group Aesthetics 1.99 (on a four-point scale). While the mean scores tended to cluster around mid-range values, the standard deviations (ranging from 0.5 to 0.8) indicate varied perspectives within the groups.

Fig. 13 (right) shows how participants perceived relationships among the four criteria. Inclusivity correlates moderately with Accessibility (0.55) and Aesthetics (0.54), implying that participants viewed physically navigable and visually appealing streets as inclusive. Practicality and Aesthetics exhibit a weak negative correlation (−0.05), highlighting that participants did not necessarily associate functional features—such as ramps or clear signage—with visually appealing design. These correlations reflect the complexity of balancing functionality, safety, and visual quality in public spaces.

To explore demographic differences in inclusivity perceptions, we examined participants' evaluations of the same 60 data points. Fig. 14 presents a boxplot showing mean and median inclusivity ratings by group. Participants identifying as elderly males, young females, or individuals with mobility impairments provided lower median scores, which aligns with interview discussions mentioning safety concerns at night, threats from cyclist traffic, and barriers posed by narrow sidewalks, limited ramps, or winter conditions. In contrast, younger males and LGBTQIA2+ participants recorded higher inclusivity scores, often referencing vibrant cultural corridors such as Avenue Laurier, where evening entertainment and mixed-use development enhanced their sense of welcome. These patterns highlight how diverse needs and symbolic cues, such as entertainment venues or supportive cultural markers, shape inclusivity perceptions (Anttiroiko & De Jong, 2020; Costanza-Chock, 2020; Rinaldi et al., 2020).

### 4.2. Model predictions

We applied the Street Review model—trained on criteria derived from semi-directed interviews, focus group ratings, and segmented features from 15,000 images (60 data points, each represented by 250 images)—to 45,000 Mapillary images of Montréal. Fig. 13 (left) illustrates the correlations among Inclusivity, Accessibility, Practicality, and Aesthetics as determined by the AI model. Accessibility and Practicality show a strong correlation (0.73), indicating that the algorithm often associates practical value with features like walkable surfaces or wide sidewalks. Inclusivity correlates more closely with Aesthetics (0.64) than with Accessibility (0.51), suggesting the model prioritizes visual qualities over physical access when estimating inclusivity. This emphasis differs somewhat from participant assessments, where Inclusivity and Accessibility were more strongly linked (0.55). Additionally, the negative correlation between Aesthetics and Accessibility/Practicality (−0.13) reflects both the model's and participants' tendency to perceive functional features (e.g., ramps, signage) as not necessarily contributing to visual appeal.

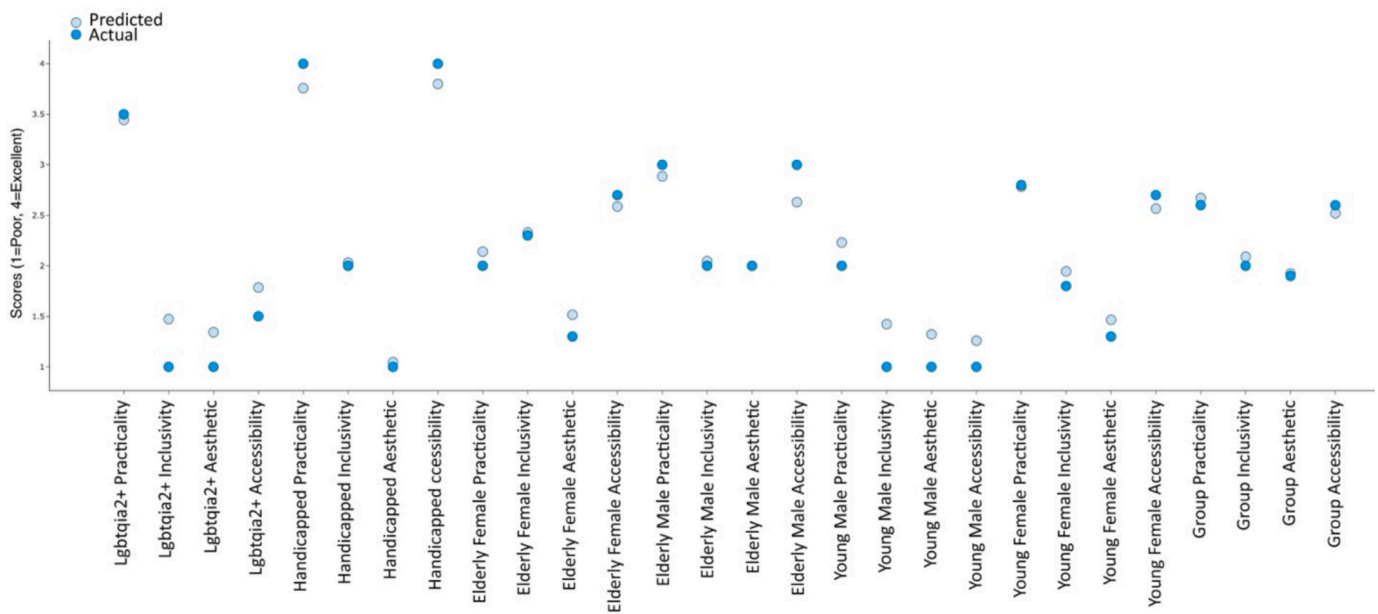

**Fig. 9.** Distribution of actual vs. predicted values for a randomly selected data point in the test set. The figure presents a comparative analysis of the model's inclusivity predictions and participant scores across demographic groups and evaluation criteria, showing areas of agreement and difference.

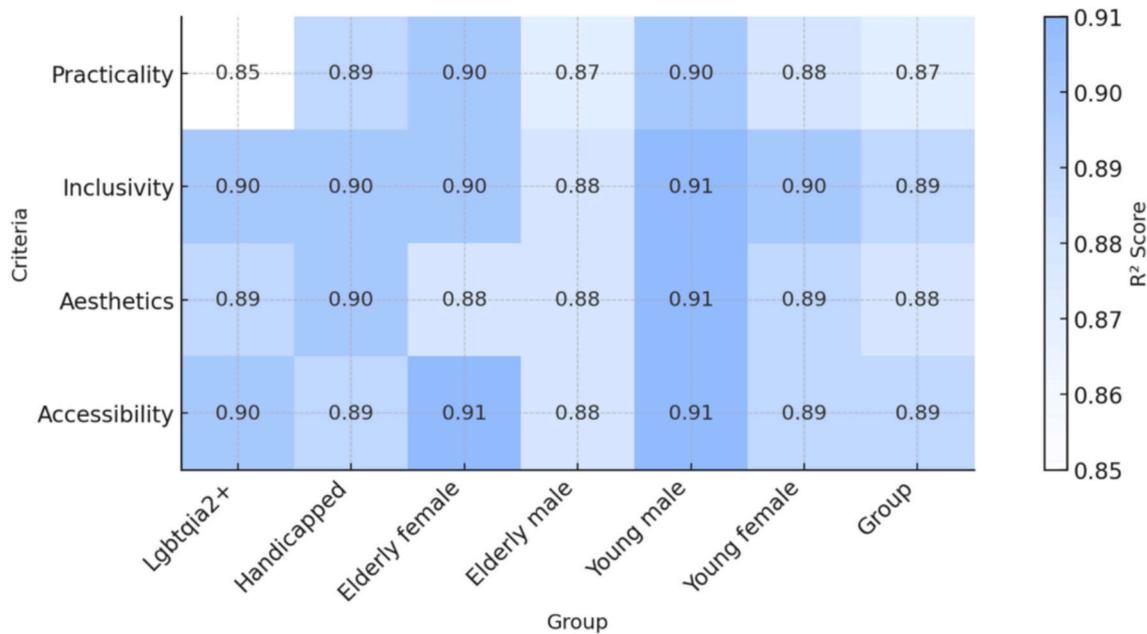

**Fig. 10.** R² scores across 28 group-criterion combinations, showing model performance variation by demographic subgroup and evaluation criterion.

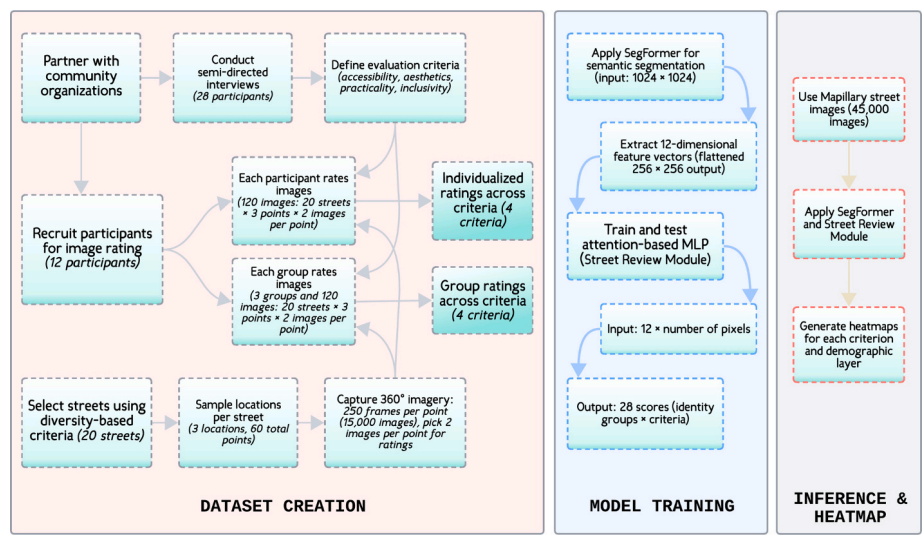

**Fig. 11.** Overview of the framework combining community input, image ratings, and model training to predict and map scores across four evaluation criteria using street-level imagery.

Fig. 15 presents the model's predictions, leveraging Mapillary's crowdsourced approach to provide extensive spatial coverage. However, as shown in Fig. 16, the quality of some images—affected by poor lighting, blurring, and distorted angles—posed challenges for accurate feature extraction. To preserve the dataset's diversity and ensure representation of various urban contexts, we opted not to filter out these lower-quality images. This decision, while broadening the dataset's scope, led to reduced accuracy in identifying features such as sidewalks, greenery, and signage, resulting in occasional inconsistencies in predicted scores, particularly in peripheral neighborhoods where such issues were more prevalent.

### 4.3. Disaggregated results

Interview data show that elderly women ($n = 13$) often emphasize continuous sidewalks, adequate lighting, and seating. Their perceived link between Inclusivity and Accessibility was moderately high (0.55), aligning with the model's emphasis on sidewalk maintenance and coverage. Still, finer details—such as tactile paving or curb

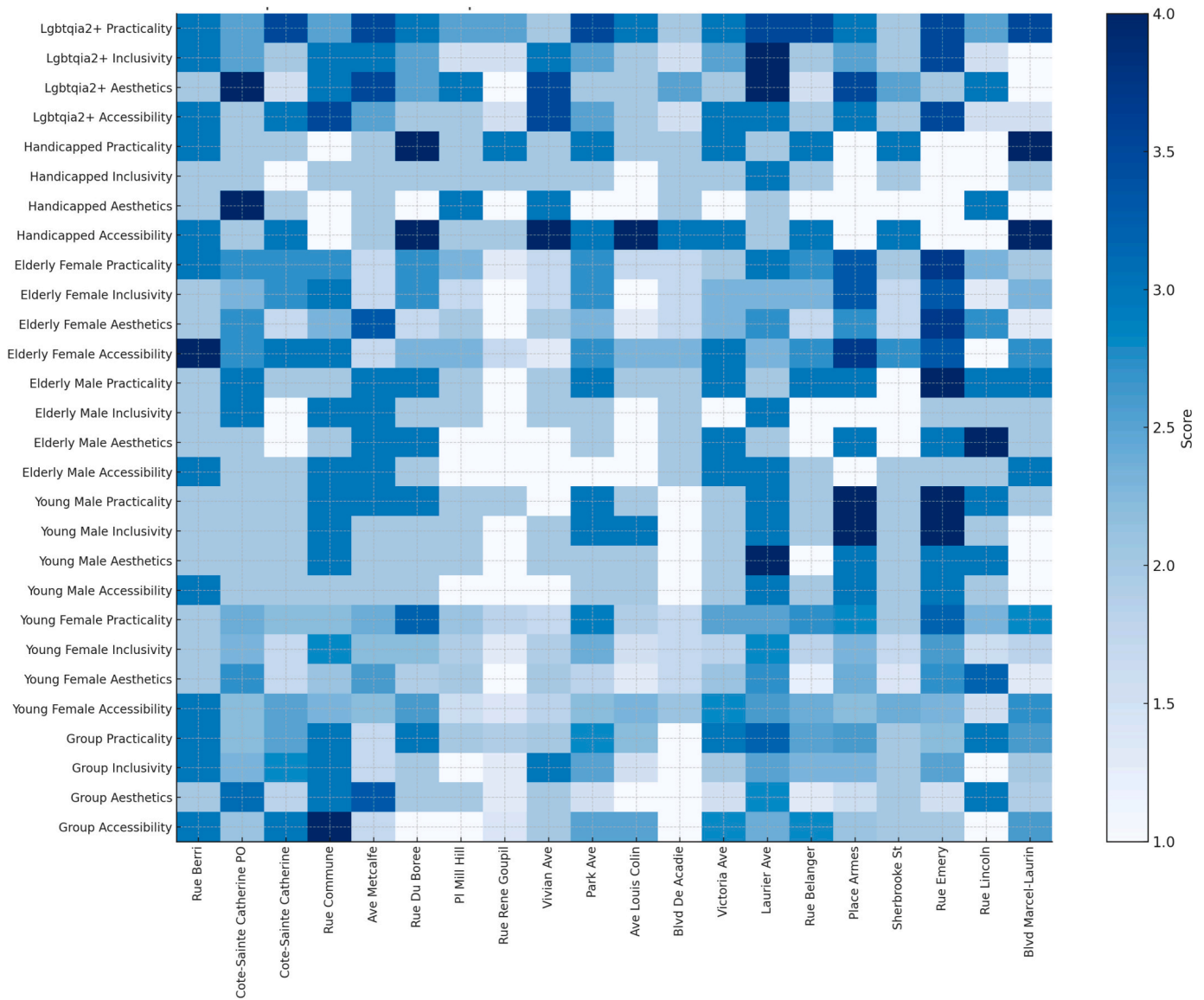

**Fig. 12.** Matrix representation of aggregated ratings for 20 Montreal streets across various dimensions, evaluated by a diverse group of participants (Mushkani & Koseki, 2025).

quality—remained difficult to detect, suggesting a need for more refined segmentation methods.

LGBTQIA2+ participants ($n = 5$) reported a moderate average inclusivity score (2.3). They noted that sporadic symbols of acceptance and lighting conditions are significant but may be overlooked by automated image analysis. This challenge aligns with the model's moderate Inclusivity–Accessibility correlation (0.51), revealing partial alignment with participant feedback yet room for further model refinement.

Participants with mobility impairments ($n = 2$) assigned consistently low scores to neighborhoods with narrow or uneven sidewalks, reflecting an average Inclusivity rating of about 1.8. Fig. 17 illustrates how the model's emphasis on mobility-related features yields lower inclusivity predictions for peripheral neighborhoods lacking robust pedestrian infrastructure. This contrast between well-equipped central areas and outlying regions highlights disparities in urban design interventions (Anttiroiko & De Jong, 2020; Low, 2020; Pettas, 2019). Fig. 18 builds on this by showing citywide spatial patterns in Montréal based on group evaluation, where higher inclusivity is concentrated in central districts and lower scores extend across the periphery.

The bottom panels of Figs. 17 and 18 contrast representative streetscapes with low and high Inclusivity. Low-scoring sites tend to lack pedestrian amenities, provide narrow sidewalks, and have limited greenery; high-scoring streets include wider walkways, more greenery, seating, and public art. These comparisons reinforce the significance of design elements in fostering or inhibiting a sense of welcome (Anttiroiko & De Jong, 2020; Armstrong & Greene, 2022).

*4.4. Citywide heatmap analysis*

We applied Street Review citywide, generating neighborhood-level inclusivity predictions. Fig. 18 reveals that central districts—such as Ville-Marie, Outremont, and areas around Mount Royal and Parc La Fontaine—scored higher, consistent with pedestrian-friendly investments and frequent maintenance. Peripheral neighborhoods show lower scores, mirroring participant accounts of infrastructure gaps (Margier, 2013). Overall, most public spaces exhibit mid-range Inclusivity across groups. For additional maps by criterion and demographic group, see https://github.com/rsdmu/streetreview.

Fig. 19 extends this analysis by comparing evaluation scores for the same image across identity groups, based on the model's predictions. The figure highlights systematic demographic differences while also revealing areas of broad agreement—particularly around aesthetic

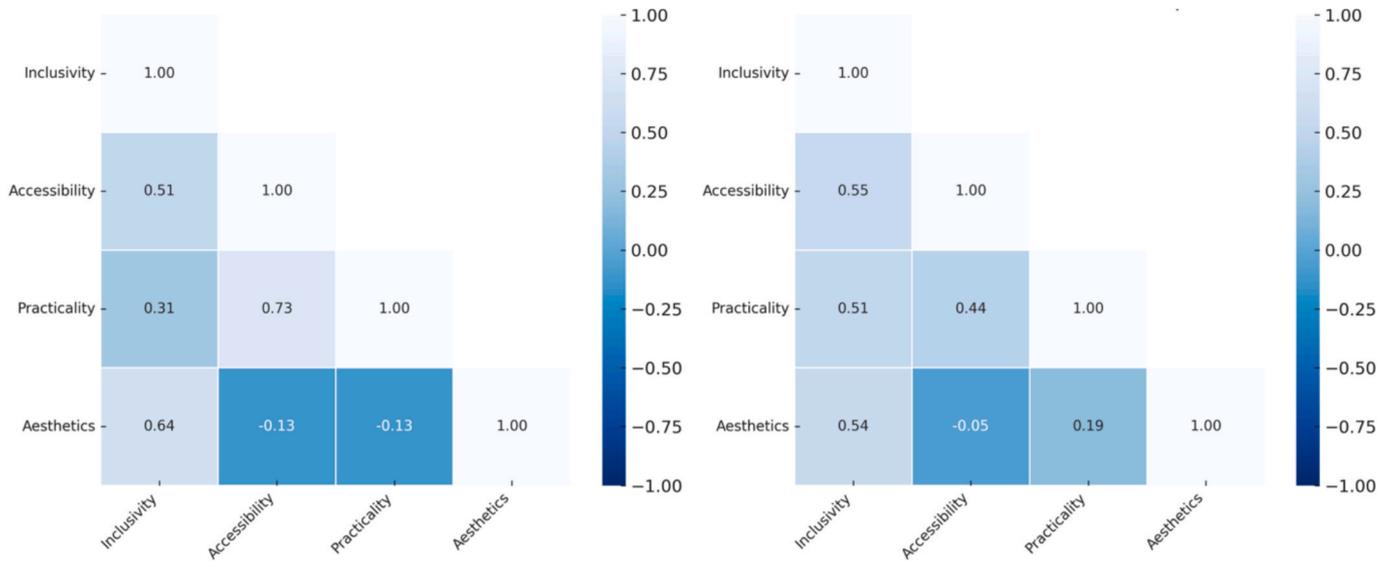

**Fig. 13.** Correlation between different criteria – Participant evaluations (right) and model predictions (left).

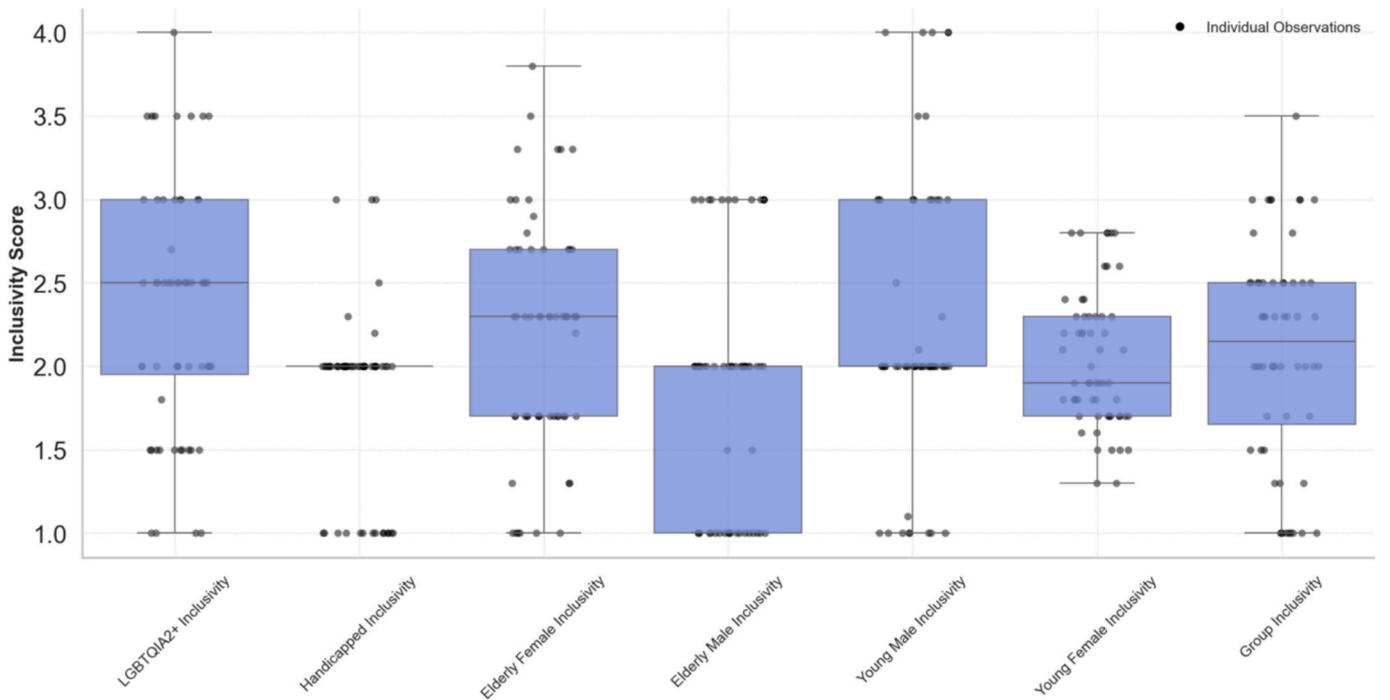

**Fig. 14.** Boxplot analysis of inclusivity ratings of 60 datapoints (20 streets) across various demographic groups, highlighting differences in perceived inclusivity.

qualities.

Overall, our findings integrate focus group assessments with AI-based segmentation and prediction to demonstrate how sidewalks, greenery, signage, seating, and cultural markers influence perceived inclusivity. While participants linked inclusive environments closely with accessibility, the Street Review model placed somewhat more weight on aesthetic factors. These insights, along with disaggregated results, highlight differences across demographic groups and the importance of refining both data and algorithms to capture complex, intersectional experiences. Despite constraints posed by uneven Mapillary coverage and image quality, the combined participatory and computational method provides an adaptable framework for identifying disparities and guiding investments in more inclusive streetscapes (Carnemolla et al., 2021; Wang et al., 2022).

## 5. Discussion

The findings indicate that intersectionality is critical when evaluating how city streets accommodate diverse groups. Variations in how participants rated accessibility, inclusivity, practicality, and aesthetics highlight the need for approaches that address distinct concerns of elderly women, LGBTQIA+ individuals, or those with disabilities (Crenshaw, 1989; Low, 2020). The Street Review model captured some of these differences through participant-generated labels and iterative calibration sessions, yet complexities remain. For example, the algorithm struggled with context-specific markers of acceptance, such as culturally significant symbols (Barocas et al., 2022; Gebru et al., 2021).

Thematic analysis of interview transcripts provided further granularity on how groups prioritize or interpret key criteria. For example,

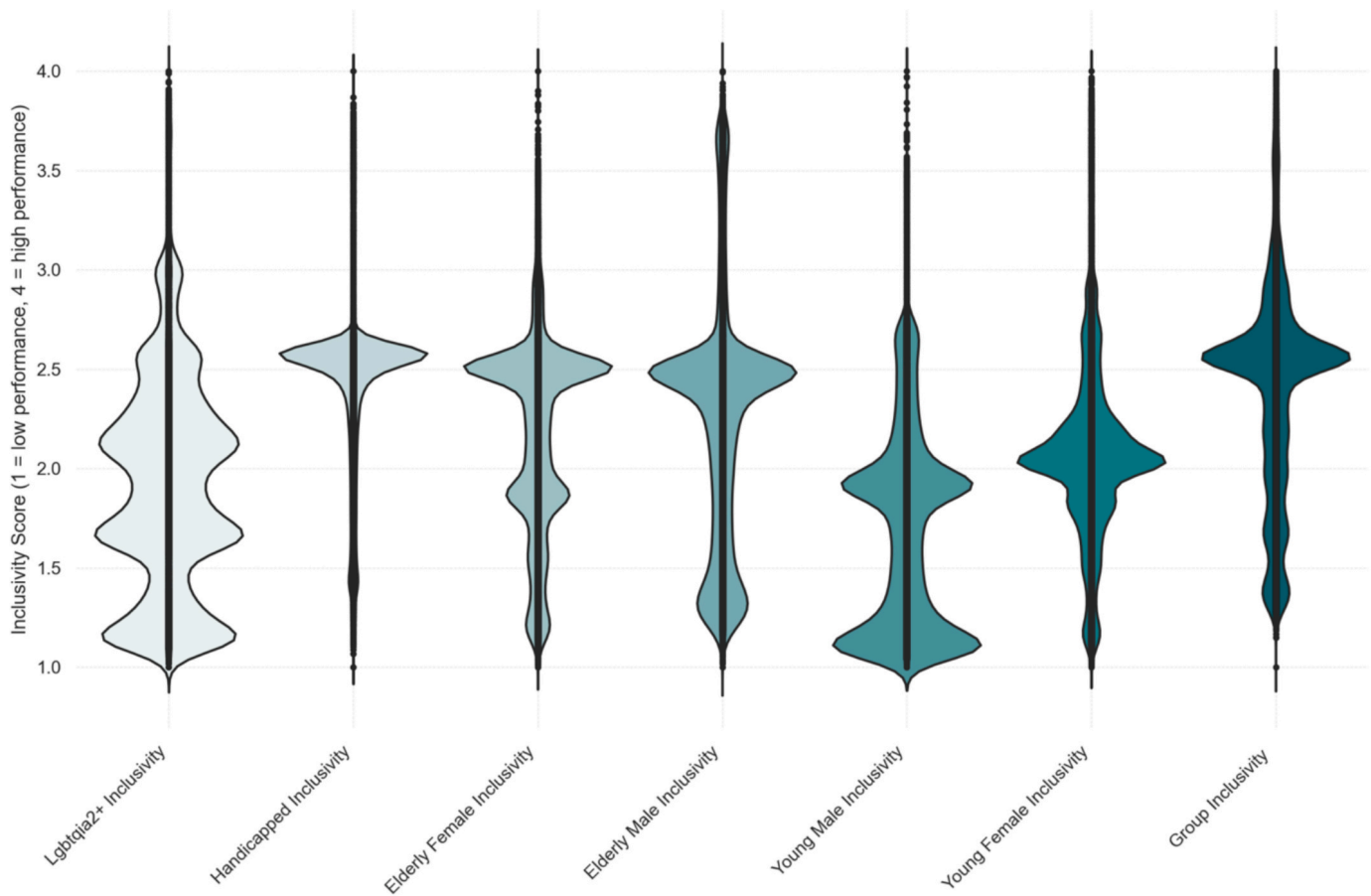

**Fig. 15.** Violin plots of inclusivity ratings for 45,000 Mapillary images, showing distribution across demographic groups.

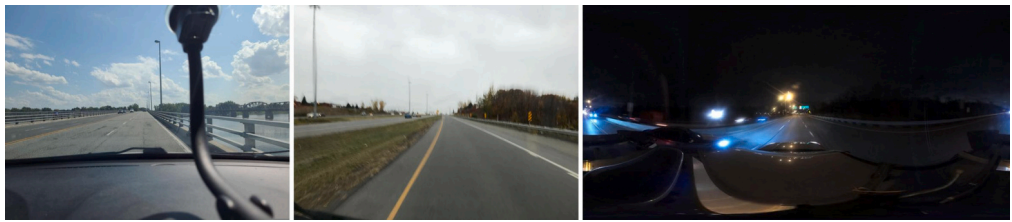

**Fig. 16.** Limitations of Mapillary. Dark, blurry, and distorted captures that constrain the model's capacity to extract accurate features and infer inclusivity.

while accessibility was a recurring concern, elderly participants often described it in terms of physical rest points and sidewalk continuity, whereas younger, able-bodied individuals focused on bike lanes and active transit. Mothers and women frequently prioritized amenities that fostered social interaction and caregiving (e.g., benches arranged for conversation), while mobility-impaired participants stressed the need for step-free design and wide sidewalks. Aesthetics, though valued by most, was often intertwined with comfort (shade, greenery, cleanliness) for some, and vibrancy or street activity for others. Experiences of safety diverged, with some groups associating police presence with reassurance, and others citing it as a source of exclusion or discrimination. Thematic matrices (Table 1) demonstrate how demographic background shaped perceptions of inclusivity, safety, and practicality, reinforcing the need for participatory, intersectional design in urban research.

Focus group discussions confirmed that co-production helps mitigate biases in model development. Inviting participants to label images, refine model outputs, and address discrepancies between predictions and personal experiences reduced some potential sources of bias. However, essentializing broad categories (e.g., "older adult," "religious minority") can oversimplify the wide spectrum of experiences within each group (Costanza-Chock, 2020). This study suggests that while broad categories can function as initial heuristics, more granular demographic layers may be required to adequately capture nuanced user experiences.

The Street Review method offers a scalable approach to producing detailed assessments of streetscape inclusivity. Urban planners and policymakers may incorporate these findings into neighborhood revitalization strategies, pedestrian master plans, or corridor studies. The observed correlation between inclusivity and accessibility suggests that investments in sidewalk infrastructure, crosswalk design, and curb cuts could enhance perceptions of inclusivity. Concurrently, aesthetic improvements, such as landscaping or façade enhancements, may further encourage a sense of welcome (Chen et al., 2024; Fan et al., 2023; Huang et al., 2023).

Municipal offices focusing on tourism and short-term visitors might also benefit from these insights. Visitors unfamiliar with local norms could encounter barriers that do not affect long-term residents (Li et al., 2022; Stark & Meschik, 2018). Clear signage, consistent wayfinding, and interactive public art might reduce language-related obstacles. The method's capacity for generating group-specific evaluations enables policymakers to assess potential trade-offs among different demographic groups (Varna & Tiesdell, 2010).

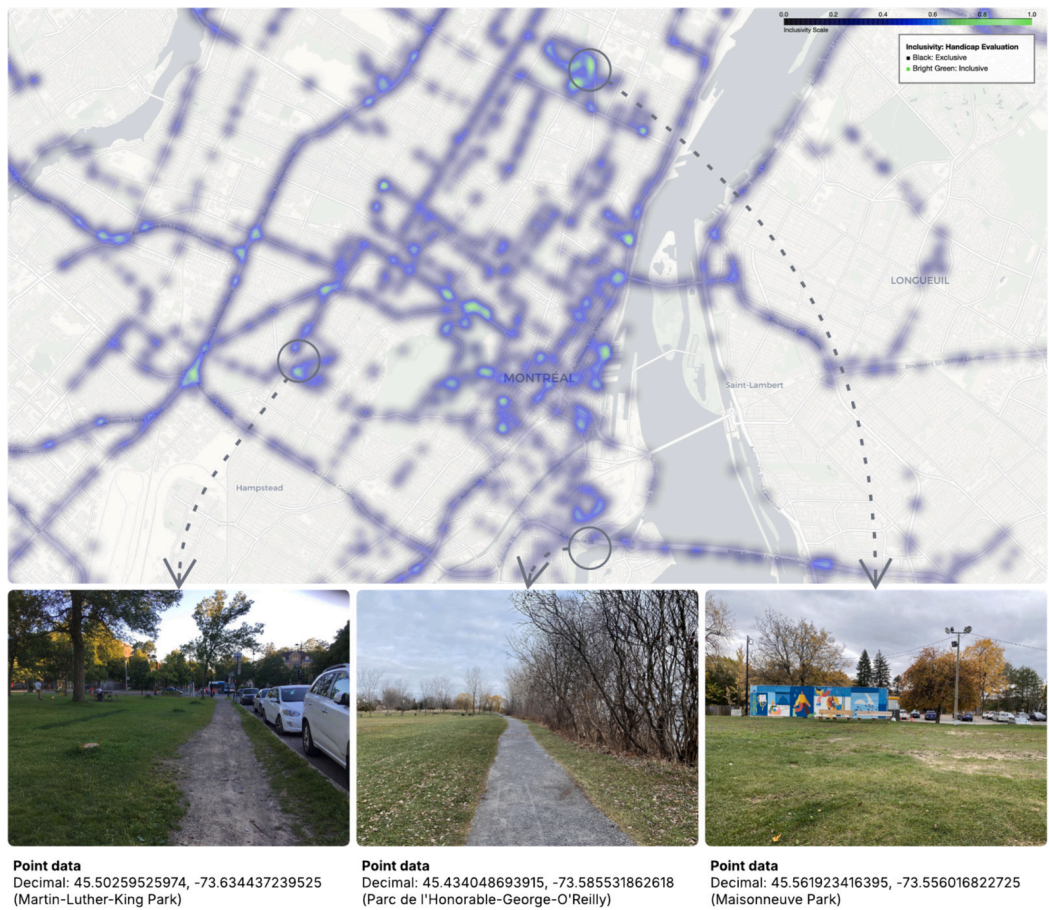

**Fig. 17.** Inclusivity heatmap based on handicap evaluation model predictions, showing how neighborhoods score on inclusivity. The heatmap was generated using the Folium library in Python, with base map data from OpenStreetMap. Bottom: Examples of streetscapes with high inclusivity scores, illustrating well-maintained sidewalks, abundant greenery, ample seating, and cultural elements, consistent with participants' perceptions of inclusive spaces.

Existing frameworks like Streetscore and Project Sidewalk use crowd-sourced or computer vision–based methods to evaluate urban design dimensions but often lack engagement with local communities' intersectional perspectives (Naik et al., 2014; Saha et al., 2019). Recent studies show that such automated approaches may diverge from local perspectives or rely on incomplete data. Kang et al. (2023) found that GeoAI-based predictions often misalign with neighborhood survey responses, while Yang et al. (2025) noted unresolved concerns around data quality and representativeness.

The Street Review framework enhances these by integrating co-production and intersectionality, allowing participants to provide nuanced, context-specific labels. While large-scale models like those developed by Ogawa et al. (2024) and Cui et al. (2023) capture patterns across extensive datasets, they may overlook group-specific perceptions and social context. In contrast, the Street Review approach incorporates qualitative interviews and focus group inputs to generate ground-truth labels. This design addresses limitations in label quality and causal inference identified by Ito et al. (2024). This hybrid approach combines computational efficiency with a deeper understanding of safety cues, cultural markers, and group-specific accessibility concerns, offering a more comprehensive assessment of inclusivity, aesthetics, practicality, and accessibility (Birhane et al., 2022; Zicari et al., 2021).

Although quantitative metrics, such as sidewalk width or building height, are central to many urban design guidelines, they only partly reflect people's lived experiences in a space (Gehl, 2011; Whyte, 2021).

The moderate-to-low correlations between practical features and perceived inclusivity or aesthetics indicate that planners should exercise caution in relying solely on objective metrics (Low, 2020; Mehta, 2014). Qualitative insights provide critical information on cultural identity, historical context, and emotional resonance (Creswell & Creswell, 2022).

The Street Review approach demonstrates how participatory research and advanced AI can be combined to produce refined inclusivity assessments:

1. Modular co-production: Participants engaged in interviews, focus groups, data labeling, and model testing, facilitating feedback loops that improved the model's performance.
2. Multi-level image sampling: We used high-resolution local photographs for detailed labeling, then scaled the analysis citywide using Mapillary.
3. Aggregated insights and heatmap generation: The approach allows negotiative group evaluations, which are then visualized in heatmaps for broader decision-making.
4. Demographic-specific heatmaps: The framework enables feature reweighting to reflect diverse user perspectives, showing how an area may seem inclusive to one group yet alienating to another.
5. Street Review dataset: The dataset includes 15,000 images captured from 60 vantage points, with each vantage point represented by 250 images. It features ratings provided by diverse individuals, both individually and as part of group evaluations. The dataset is available on Huggingface for further model training or downstream fine-tuning.

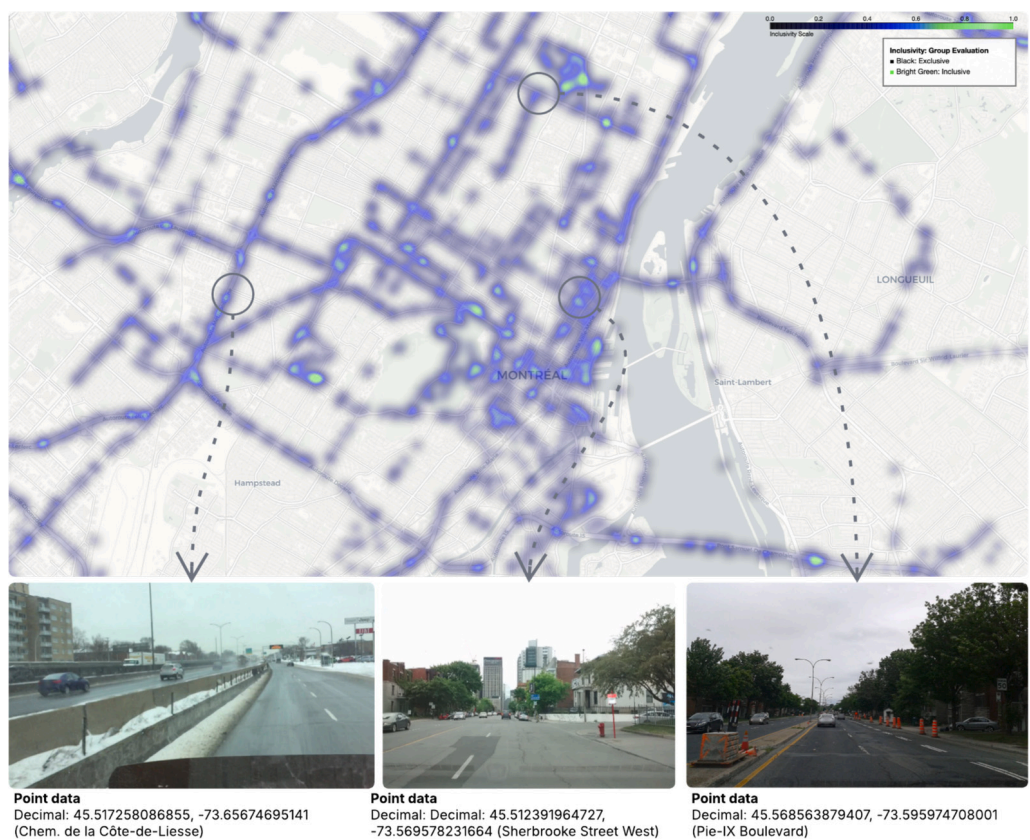

**Fig. 18.** Citywide spatial patterns in Montréal, where green areas indicate higher predicted inclusivity and darker regions indicate lower scores. The heatmap uses the Folium library in Python, with base map data from OpenStreetMap. For interactive map, please visit https://mid-spaces.github.io/landing-page/montreal_folium _heatmap_group_inclusivity.html.
Bottom: Examples of streetscapes with low inclusivity scores, featuring limited pedestrian amenities, inadequate sidewalks, and a lack of community-oriented features. (For interpretation of the references to color in this figure legend, the reader is referred to the web version of this article.)

These methodological innovations can inform future urban planning research and practice, illustrating how large-scale inclusive audits can be conducted while respecting the heterogeneity of urban communities.

## 6. Limitations

This research faced several challenges. First, co-production is resource-intensive, requiring multiple interactions with participants. Attrition occurred when some community members withdrew before finishing all modules. Second, dependence on Mapillary imagery introduced constraints related to coverage and image quality, especially in neighborhoods with limited representation. Third, the model predominantly detects observable physical elements; intangible factors like cultural cues or personal experiences of harassment are more difficult to capture through image segmentation alone. Fourth, participants were grouped into broad demographic categories (e.g., older adult, LGBTQIA+), which can mask intragroup variation.

A further limitation involves the study size, especially in the image rating and focus group phase. Only 12 participants took part in this stage. This number limited the breadth of perspectives included in the scoring of images and in group discussions, reducing the ability to capture a wide range of views on inclusivity. The relatively small participant pool, along with limited citywide data points, imposes constraints on statistical generalization. Improving these aspects may involve more extensive image collection, finer-grained demographic labels, and integration of real-time datasets, such as footfall counts or noise levels, to better reflect temporal dynamics of inclusivity.

## 7. Implications for urban policies and planning

Street Review's integration of co-production and AI-based analysis offers policymakers a structured method to identify design features that enhance perceived inclusivity. The study indicates that sidewalks, building frontages, and walls exert a greater influence on inclusivity ratings than greenery, and that group-level evaluations often yield more calibrated outcomes than individual assessments. By combining focus groups, interviews, and machine learning, the approach surfaces user-defined priorities, such as symbolic markers, sidewalk maintenance, and localized safety concerns, that may not appear in conventional guidelines. Its open-source design permits replication in smaller municipalities or under-resourced contexts, where communities can collect images to supplement or replace large-scale databases. Policymakers might use demographic-specific weightings and heatmaps to target street-level interventions, emphasizing both accessible and visually appealing environments in order to increase perceptions of inclusivity.

## 8. Conclusion

This paper introduced Street Review, a participatory methodology that combines qualitative methods with AI-based image analysis to evaluate streetscape inclusivity. Semi-directed interviews and focus groups in Montréal revealed that sidewalks, greenery, symbolic markers, seating availability, and lighting are critical to participants, though priorities differ across demographic groups. Meanwhile, the AI model primarily recognized sidewalk coverage, walls, and building features as major predictors of inclusivity, underscoring the necessity of accounting for intangible, culture-specific dimensions not always captured through image segmentation.

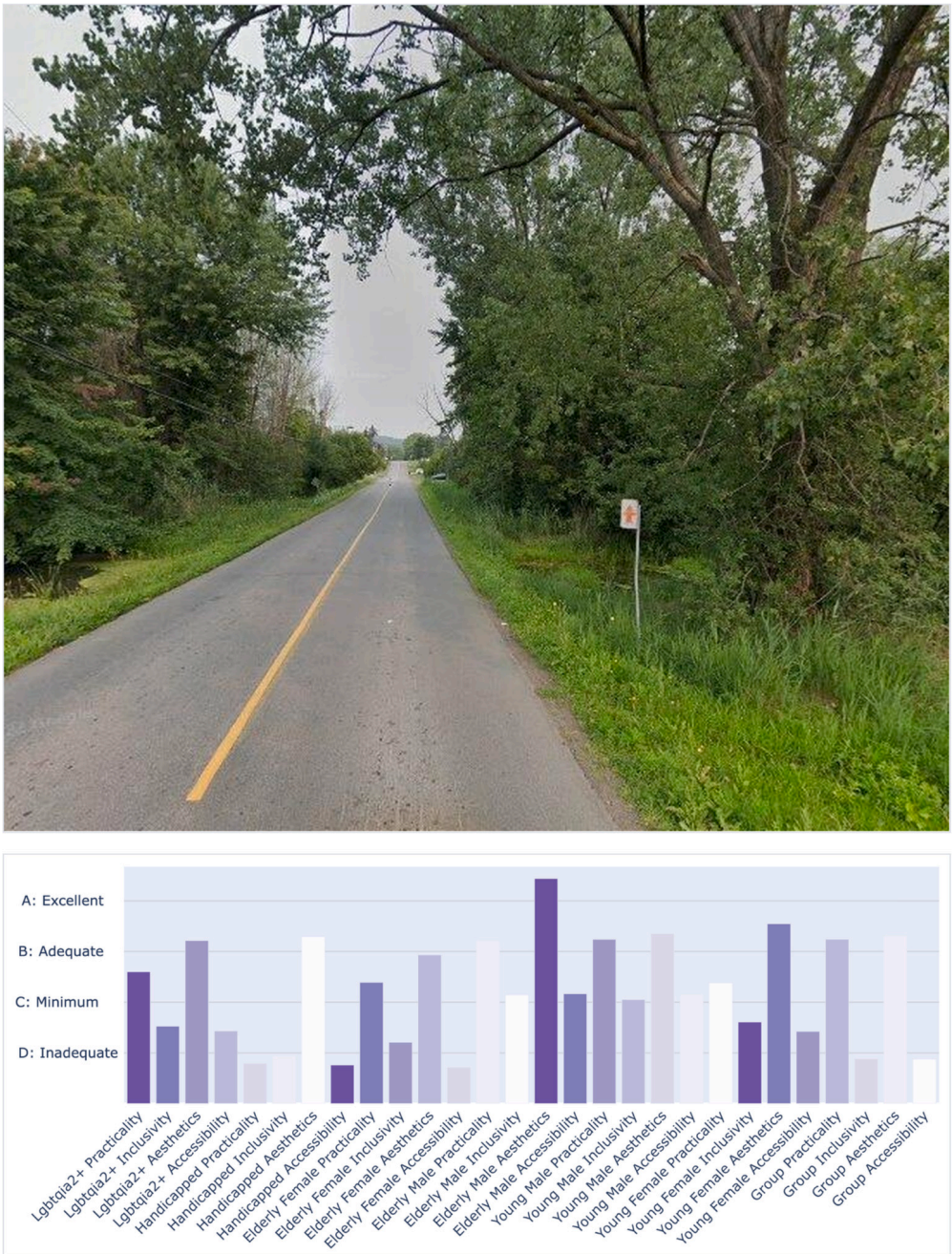

**Fig. 19.** Divergent group evaluations of a single streetscape. Top: source image; bottom: predicted grades (A = Excellent–D=Inadequate) for Inclusivity, Accessibility, Practicality, and Aesthetics across demographic-specific models.

By incorporating participant-generated labels into a multi-output regression model, we generated citywide heatmaps of inclusivity, accessibility, aesthetics, and practicality. All maps are accessible through this URL in a GitHub repository: https://github.com/rsdmu/streetreview. Both human evaluations and model predictions suggested a strong correlation between inclusivity and accessibility, although the algorithm struggled with intangible cultural elements. Demographic-specific weightings showed that a single environment could be inclusive for one group yet relatively inaccessible for another.

This approach remains limited by image quality, number of participants, coarse demographic categories, and the resource demands of co-production, yet it offers a structured path to refine future audits. Future refinements of Street Review may include:

1. Longitudinal studies: Monitoring changes in inclusivity as neighborhood conditions evolve.
2. Partnering with a larger number of participants.
3. Real-time or sensor data integration: Enhancing model predictions through footfall counts or geotagged social media posts.
4. Comparative analyses: Adapting Street Review for multiple cities to examine how cultural, climatic, or policy differences affect inclusivity.
5. Advanced symbol detection: Refining algorithms to better identify cultural or symbolic markers while maintaining ethical standards.
6. Intersectional demographic categories: Moving beyond broad labels to capture the realities of individuals with multiple overlapping identities.

In rapidly diversifying urban environments, Street Review offers planners and policymakers a practical and scalable approach to identifying exclusionary features, guiding investment, and fostering genuinely inclusive urban spaces.

## CRediT authorship contribution statement

**Rashid Mushkani:** Writing – review & editing, Writing – original draft, Visualization, Validation, Software, Methodology, Investigation, Funding acquisition, Formal analysis, Data curation, Conceptualization. **Shin Koseki:** Supervision, Resources, Project administration, Funding acquisition.

## Code and dataset availability

The source code is publicly available on GitHub at https://github.com/rsdmu/streetreview, while the dataset can be found on Hugging Face at https://huggingface.co/datasets/rsdmu/streetreview.

## Ethical statements

This study was approved by the appropriate Research Ethics Committee.

## Funding

This research was supported by the Québec Research Fund (FRQ; https://doi.org/10.69777/347989) and the Social Sciences and Humanities Research Council of Canada (Grant No. NFRFR-2021-00397).

## Declaration of competing interest

The authors declare no conflicts of interest.

## Data availability

The datasets generated during the current study are available in a Hugging Face repository.